\documentclass[preprint,12pt]{elsarticle}

\usepackage{graphicx}

\usepackage{amssymb}
\usepackage{amsmath}

\journal{Journal of Computational Physics}

\begin{document}

\begin{frontmatter}



\title{Note on the use of Yee-lattices in (semi-) implicit Particle-in-cell
codes}


\author{Andreas Kempf}
\ead{akempf@astro.uni-wuerzburg.de}
\author{Urs Ganse}
\author{Patrick Kilian}
\author{Felix Spanier}

\address{Lehrstuhl f\"ur Astronomie, Universit\"at W\"urzburg,
	Emil-Fischer-Stra{\ss}e 31, D-97074 W\"urzburg}

\begin{abstract}
	A modification of the implicit algorithm for particle-in-cell simulations
	proposed by \citet{petdav} is presented. The original lattice arrangement
	is not inherently divergence-free, possibly leading to unphysical results.
	This arrangement is replaced by a staggered mesh resulting in a reduction
	of the divergence of the magnetic field by several orders of magnitude.
\end{abstract}


\end{frontmatter}


\section{Introduction}
\label{sec:intro}
	In order to correctly reproduce physical processes in a particle-in-cell
	code, Maxwell's equations need to be solved consistently. However, the
	requirement of the magnetic field being divergence free is often violated by
	numerical algorithms leading to unphysical results
	\citep{brackbill_divergence}.  Consequently, several divergence-cleaning
	schemes have been proposed, providing a way to remove magnetic source terms
	after the fact.  Another possibility to correctly incorporate Gauss' law for
	magnetism into a PiC-code is to use the staggered mesh first proposed by
	\citet{yee} in 1966.  The special arrangement of electric and magnetic fields
	inherently conserves a zero-valued divergence \citep{taflove}, provided that
	$\nabla \cdot \vec{B} = 0$ at $t=0$. A comparison by \citet{baki} identifies
	several problems of divergence-cleaning methods in MHD and notes their
	absence when using a staggered mesh.

	\citet{petdav} proposed an implicit particle-in-cell algorithm forgoing the
	staggered mesh approach. Continuing previous work by \citet{Kilian_2011} we
	intend to use this algorithm to study particle acceleration in astrophysical
	plasmas while keeping unphysical effects to a minimum.  In this paper we
	therefore modify the scheme, incorporating the Yee lattice and effecting a
	reduction of $\nabla \cdot \vec{B}$ by several orders of magnitude.

\section{Definitions}
	The quantities from \citep{petdav} that are relevant to this paper are

	\begin{equation}
		\hat{S}_{\alpha}^{\,n+1/2} = \frac{n_{\alpha} q_{\alpha}}
		{4 \varepsilon_{0} m_{\alpha} \gamma_{\alpha}^{\,n+1/2}}
		\hat{T}_{\alpha}^{\,n+1/2}
	\end{equation}
	and
	\begin{equation} \label{eq:last}
		\delta \vec{j}_{\alpha}^{\,n+1/2} = 
			\frac{n_{\alpha} q_{\alpha}}
			{2 m_{\alpha} \gamma_{\alpha}^{\,n+1/2}}
			\left(
				\vec{p}_{\alpha}^{\,n}
				+ \hat{T}_{\alpha}^{\,n+1/2}
				\left(
					\vec{p}_{\alpha}^{\,n}
					\times \Delta \vec{\Omega}_{\alpha}^{\,n+1/2}
				\right)
			\right)
			\mbox.
	\end{equation}
	The tensor $\hat{T}$ (with indices suppressed for brevity) is defined as
	\begin{equation} \label{eq:magrot}
		\hat{T} =
		\frac{1}{1 + |\Delta
		\vec{\Omega}|^2}
			\begin{bmatrix}
				1 + \Delta \Omega_{\mathrm{x}}^2

					& \Delta \Omega_{\mathrm{x}}
						\Delta \Omega_{\mathrm{y}}
						+ \Delta \Omega_{\mathrm{z}}

					& \Delta \Omega_{\mathrm{x}}
						\Delta \Omega_{\mathrm{z}}
						- \Delta \Omega_{\mathrm{y}} \\
				\Delta \Omega_{\mathrm{x}}
					\Delta \Omega_{\mathrm{y}}
					- \Delta \Omega_{\mathrm{z}}

					& 1 + \Delta \Omega_{\mathrm{y}}^2

					& \Delta \Omega_{\mathrm{y}}
					\Delta \Omega_{\mathrm{z}}
					+ \Delta \Omega_{\mathrm{x}} \\
					\Delta \Omega_{\mathrm{x}}
					\Delta \Omega_{\mathrm{z}}
					+ \Delta \Omega_{\mathrm{y}}

					& \Delta \Omega_{\mathrm{y}}
					\Delta \Omega_{\mathrm{z}}
					- \Delta \Omega_{\mathrm{x}}

					& 1 + \Delta \Omega_{\mathrm{z}}^2
			\end{bmatrix}
	\end{equation}
	with
	\begin{equation} \label{eq:omega}
		\Delta \vec{\Omega}_{\alpha}^{\,n+1/2} =
		\frac{q_{\alpha}\vec{B}_{\alpha}^{\,n+1/2}}
			{m_{\alpha}
		\gamma_{\alpha}^{\,n+1/2}}
		\frac{\Delta t} {2}
		\mbox.
	\end{equation}
	The quantities $q_{\alpha}$, $m_{\alpha}$, $n_{\alpha}$ are the charge, mass,
	and number density of (computational) particle $\alpha$.
	$\gamma_{\alpha}^{n+1/2}$ is the particle's relativistic gamma factor and
	$\vec{B}_{\alpha}^{n+1/2}$ its local magnetic field at time $n+1/2$.
	$\vec{p}_{\alpha}^{\,n}$ is the momentum of particle $\alpha$ at time $n$.

	The deposition of $\hat{S}$ and $\delta \vec{j}$ on the grid and the
	interpolation of $\vec{E}$ and $\vec{B}$ to the particle position is achieved
	via a standard weighting function. Our algorithm makes use of the triangular
	shaped cloud (TSC) scheme.

\section{The modified lattice arrangement} \label{sec:modification}
	The original algorithm by \citet{petdav} stores electric fields on grid
	nodes and magnetic fields in the cell center. The vector quantity $\delta
	\vec{j}$ and the tensor quantity $\hat{S}$ are deposited on grid
	nodes, as well. Since the electric field is updated according to
	\begin{equation} \label{eq:E-feld}
	\left(
	\hat{\mathrm{I}}+\hat{S}^{\,n+1/2}
	\right)
	\vec{E}^{\,n+1} = 
	\left(
	\hat{\mathrm{I}}-\hat{S}^{\,n+1/2}
	\right)
	\vec{E}^{\,n}
	+ \frac{\Delta t}{\varepsilon_{0}}
	\left(
	\vec{\nabla} \times
	\vec{H}^{\,n+1/2} - \delta \vec{j}^{\,n+1/2}
	\right),
	\end{equation}
	and all required quantities are defined on grid nodes, this equation can be
	solved locally for $\vec{E}^{\,n+1}$.

	Our approach keeps the original field layout by \citet{yee} with the
	components of $\delta \vec{j}$ stored like the corresponding components of
	the electric field. $\hat{S}$ is stored on grid nodes and interpolated
	linearly for each component of the electric field to be calculated. When
	calculating the new value for $E_{\mathrm{x}}^{\,i+1/2, j, k}$, $\hat{S}$ is
	taken to be $(\hat{S}^{\,i, j, k}+\hat{S}^{\,i+1, j, k})/2$, for
	$E_{\mathrm{y}}^{\,i, j+1/2, k}$ it is
	$(\hat{S}^{\,i, j, k}+\hat{S}^{\,i, j+1, k})/2$ and for
	$E_{\mathrm{z}}^{\,i, j, k+1/2}$ it is
	$(\hat{S}^{\,i, j, k}+\hat{S}^{\,i, j, k+1})/2$.

	Since $\hat{S}$ is not a diagonal tensor, all the components of $\delta \vec{
	j}$, $\nabla \times \vec{B}$ and $\vec{E}^{\, n}$ need to be known at the
	same point as the component of $\vec{E}^{\,n+1}$ to be calculated, as well.
	These three quantities can be interpolated the same way.

	For $E_{\mathrm{x}}^{\,n+1}$:
	\begin{align}
		&A_{\mathrm{x}}^{\,i+1/2, j, k} = A_{\mathrm{x}}^{\,i+1/2, j, k} \\
		&A_{\mathrm{y}}^{\,i+1/2, j, k} = \left(A_{\mathrm{y}}^{\,i, j+1/2, k}
			+ A_{\mathrm{y}}^{\,i, j-1/2, k}
			+ A_{\mathrm{y}}^{\,i+1, j+1/2, k}
			+ A_{\mathrm{y}}^{\,i+1, j-1/2, k}\right)/4 \\
		&A_{\mathrm{z}}^{\,i+1/2, j, k} = \left(A_{\mathrm{z}}^{\,i, j, k+1/2}
			+ A_{\mathrm{z}}^{\,i, j, k-1/2}
			+ A_{\mathrm{z}}^{\,i+1, j, k+1/2}
			+ A_{\mathrm{z}}^{\,i+1, j, k-1/2}\right)/4
	\end{align}

	For $E_{\mathrm{y}}^{\,n+1}$:
	\begin{align}
		&A_{\mathrm{x}}^{\,i, j+1/2, k} = \left(A_{\mathrm{x}}^{\,i+1/2, j, k}
			+ A_{\mathrm{x}}^{\,i+1/2, j+1, k}
			+ A_{\mathrm{x}}^{\,i-1/2, j, k}
			+ A_{\mathrm{x}}^{\,i-1/2, j+1, k}\right)/4 \\
		&A_{\mathrm{y}}^{\,i, j+1/2, k} = A_{\mathrm{y}}^{\,i, j+1/2, k} \\
		&A_{\mathrm{z}}^{\,i, j+1/2, k} = \left(A_{\mathrm{z}}^{\,i, j, k+1/2}
			+ A_{\mathrm{z}}^{\,i, j+1, k+1/2}
			+ A_{\mathrm{z}}^{\,1, j, k-1/2}
			+ A_{\mathrm{z}}^{\,i, j+1, k-1/2}\right)/4
	\end{align}

	For $E_{\mathrm{z}}^{\,n+1}$:
	\begin{align}
		&A_{\mathrm{x}}^{\,i, j, k+1/2} = \left(A_{\mathrm{x}}^{\,i+1/2, j, k}
			+ A_{\mathrm{x}}^{\,i+1/2, j, k+1}
			+ A_{\mathrm{x}}^{\,i-1/2, j, k}
			+ A_{\mathrm{x}}^{\,i-1/2, j, k+1}\right)/4 \\
		&A_{\mathrm{y}}^{\,i, j, k+1/2} = \left(A_{\mathrm{y}}^{\,i, j+1/2, k}
			+ A_{\mathrm{y}}^{\,i, j+1/2, k+1}
			+ A_{\mathrm{y}}^{\,i, j-1/2, k}
			+ A_{\mathrm{y}}^{\,i, j-1/2, k+1}\right)/4 \\
		&A_{\mathrm{z}}^{\,i, j, k+1/2} = A_{\mathrm{z}}^{\,i, j, k+1/2}
	\end{align}

\section{Simulation setup}
	In order to compare the original lattice configuration with the Yee lattice,
	identical simulations are performed using both algorithms.

	The size of the three-dimensional lattice is $64 \times 64 \times 64$ with
	periodic boundary conditions. There are 25 electrons and 25 protons in each
	cell. Each simulation is running for 4000 timesteps. The plasma is chosen
	to be thermal, meaning that the velocity components for each particle are
	independent and follow a Gaussian distribution of width $v_{\,\mathrm{th}}$.
	The protons and electrons are in thermal equilibrium so that the width of
	the Gaussian distribution of the protons is equal to $v_{\,\mathrm{th}}$
	divided by the square root of the  mass ratio $\sqrt{m_{\mathrm{p}} /
	m_{\mathrm{e}}}$.

	The physical parameters are listed in table \ref{tab:parameters}.

	\begin{table}
		\centering
		\begin{tabular}{l|c|r}
			electron plasma frequency & $\omega_{\mathrm{pe}}$ & $2.0\cdot10^{8}\,
			\mathrm{rad/s}$ \\
			length of timestep & $\Delta t$ & $4.1 \cdot 10^{-10}\,\mathrm{s}$ \\
			Debye length & $\lambda_{\mathrm{D}}$ & $7.5\,\mathrm{cm}$ \\
			cell edge length & $\Delta x$ & $21\,\mathrm{cm}$ \\
			mass ratio& $m_{\mathrm{p}} / m_{\mathrm{e}}$ & 42
		\end{tabular}
		\caption{Parameters of the thermal plasma.}
		\label{tab:parameters}
	\end{table}

\section{Results}
	When directly comparing $\nabla \cdot \vec{B}$ at the end of the two
	simulation runs, a significant difference is manifest. The ratio of the total
	divergence of both simulations at timestep 4000 is
	\begin{equation}
		\frac{\sum_{ijk} |\nabla \cdot \vec{B}_{\mathrm{Yee}}|}
		{\sum_{ijk} |\nabla \cdot \vec{B}_{\mathrm{original}}|}
		= 7.6 \cdot 10^{-15}
	\end{equation}

\begin{figure}
		\centering
		\includegraphics[width=0.48\textwidth]{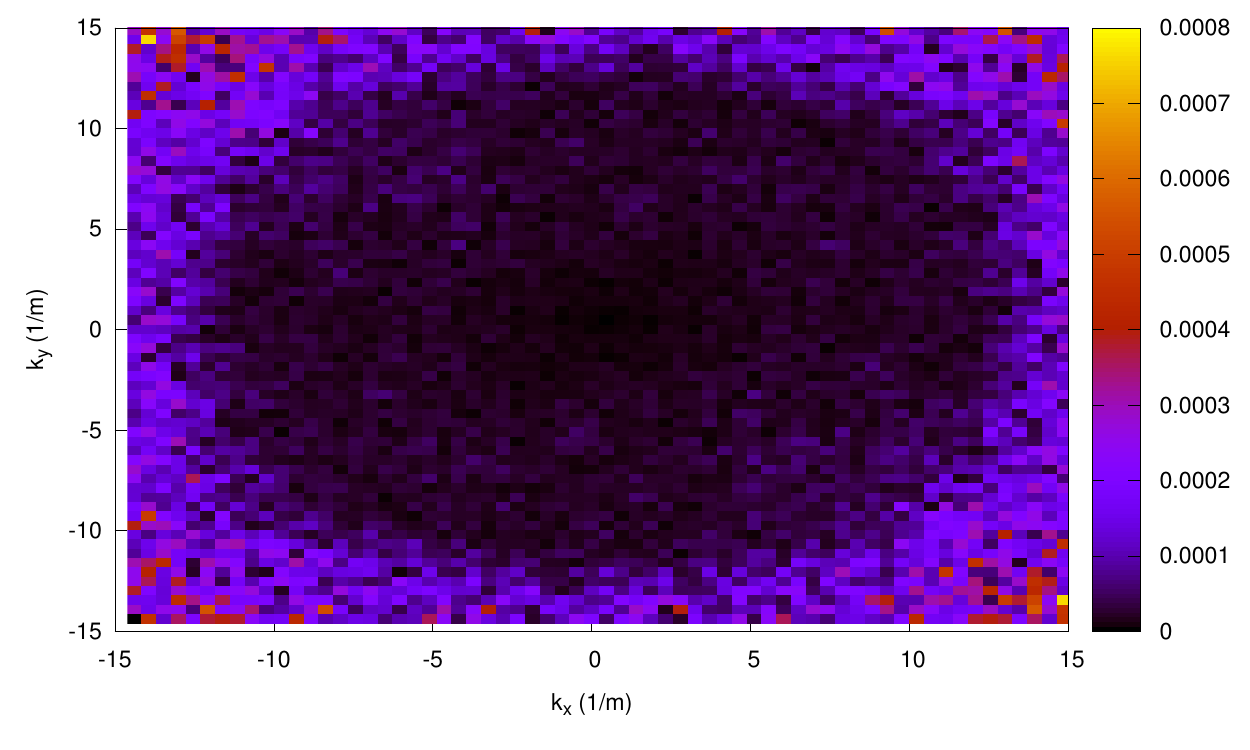}
		\includegraphics[width=0.48\textwidth]{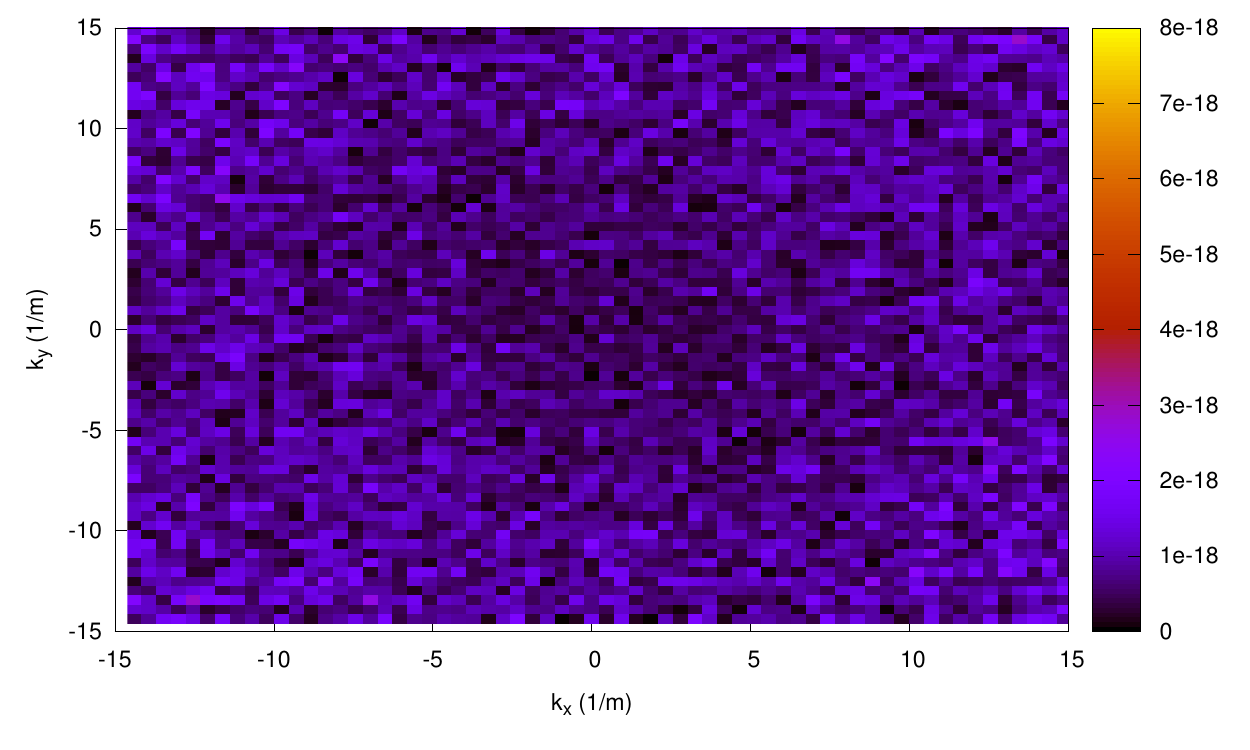}
		\caption{Spectral plot of the distribution of $\nabla \cdot \vec{B}$ in x-
			and y-direction for the original lattice arrangement (left) and the Yee
			arrangement (right).}
		\label{fig:div}
	\end{figure}

	Fourier transforming $\nabla \cdot \vec{B}$ at timestep 4000 in space and
	plotting the absolute values for a slice along the z-axis yields figure
	\ref{fig:div}. As can be seen, the original arrangement introduces
	high-amplitude patterns at short wavelengths. The Yee-arrangement mostly
	produces low-amplitude noise, as desired.

\begin{figure}
		\centering
		\includegraphics[width=0.48\textwidth]{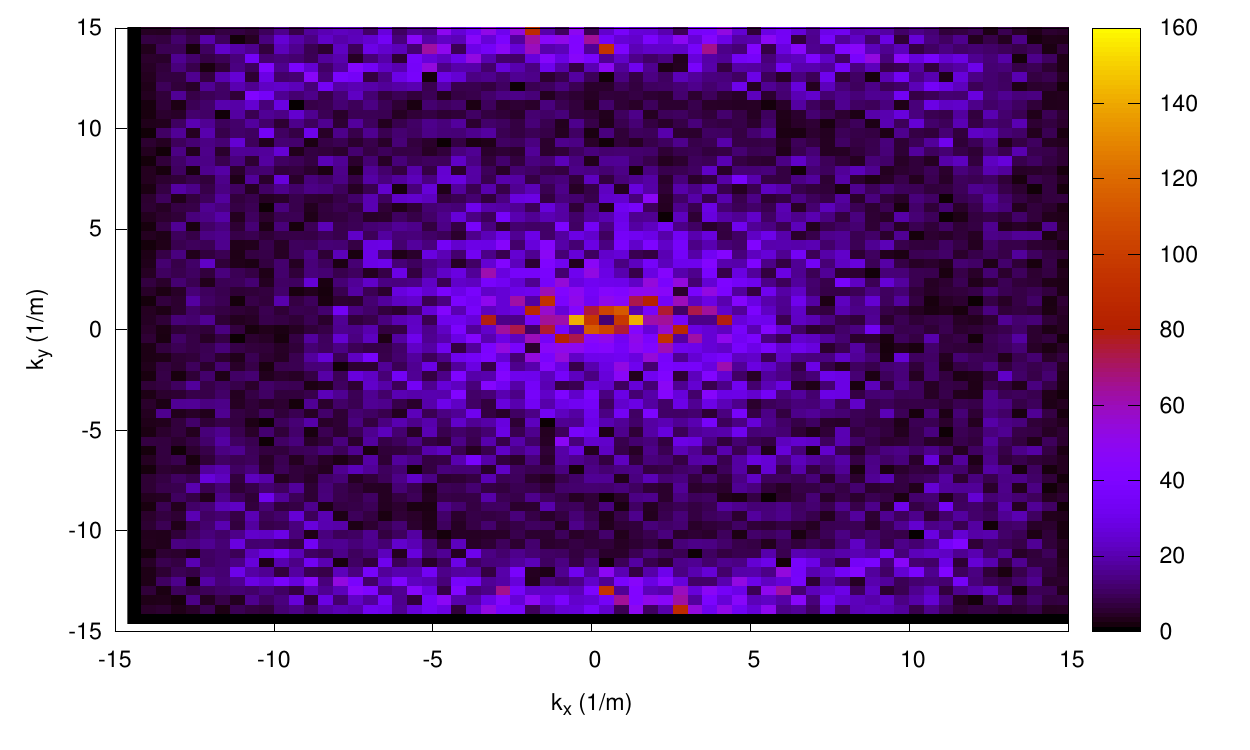}
		\includegraphics[width=0.48\textwidth]{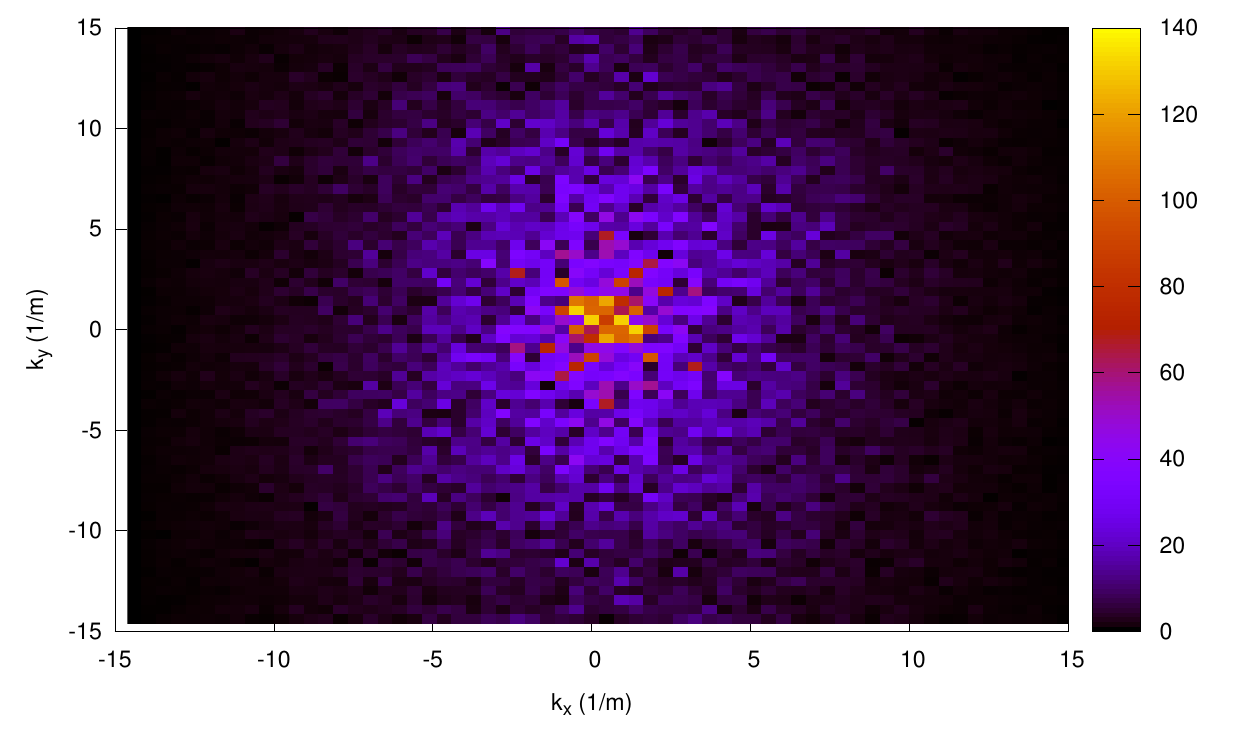}
		\caption{Spectral plot of the distribution of $B_{\mathrm{y}}$ in x- and
		y-direction for the original lattice arrangement (left) and the Yee
		arrangement (right).}
		\label{fig:magfield}
	\end{figure}

	Furthermore, plots of the y-component of the magnetic field, prepared the
	same way, are shown in figure \ref{fig:magfield}. The original lattice
	arrangement introduces unphysical behavior at short wavelengths which is not
	present when using a Yee-lattice.

\begin{figure}
		\centering
		\includegraphics[width=0.75\textwidth]{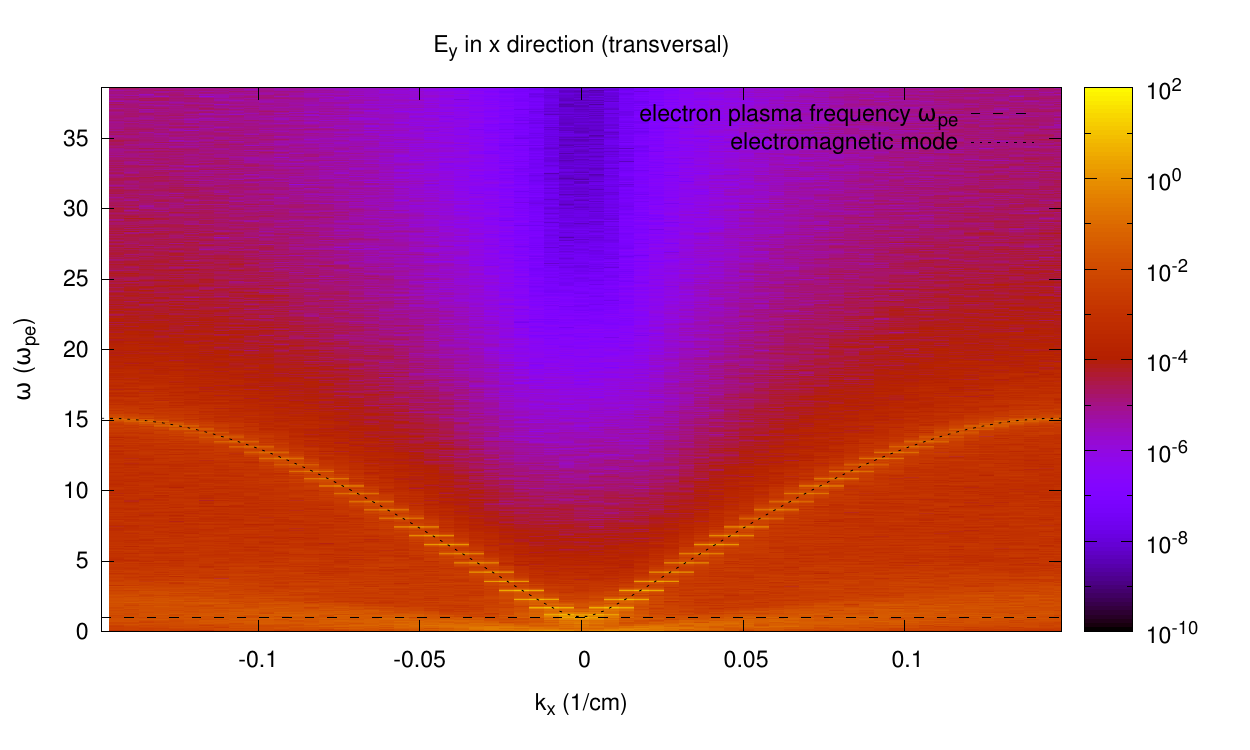}
		\caption{Plot of the dispersion relation of $E_{\mathrm{y}}$ along the
		$\mathrm{x}$-axis. The expected dispersion of the electromagnetic wave,
		incorporating numerical effects, is also shown.}
		\label{fig:dispersion}
	\end{figure}

	A dispersion plot of the $\mathrm{y}$-component of the electric field along
	the $\mathrm{x}$-axis for the simulation using the Yee-lattice is obtained as
	follows. For each timestep the $E_{\mathrm{y}}$ values are summed up for
	slices perpendicular to the $\mathrm{x}$-axis. A Fourier transformation in
	space and time results in a $\omega$-$k_{\mathrm{x}}$ relation, depicted in
	figure \ref{fig:dispersion}. In addition, the dispersion of the
	electromagnetic wave as predicted by theoretical calculations is shown
	(incorporating effects from finite $\Delta t$ and $\Delta x$). As can be
	seen, the modified code correctly describes wave dispersion in a thermal
	plasma indicating that the alterations do not introduce unphysical effects.
	Likewise, further tests of more complicated simulation setups do not show
	unexpected behavior resulting from our changes.

\begin{figure}
		\centering
		\includegraphics[width=0.48\textwidth]{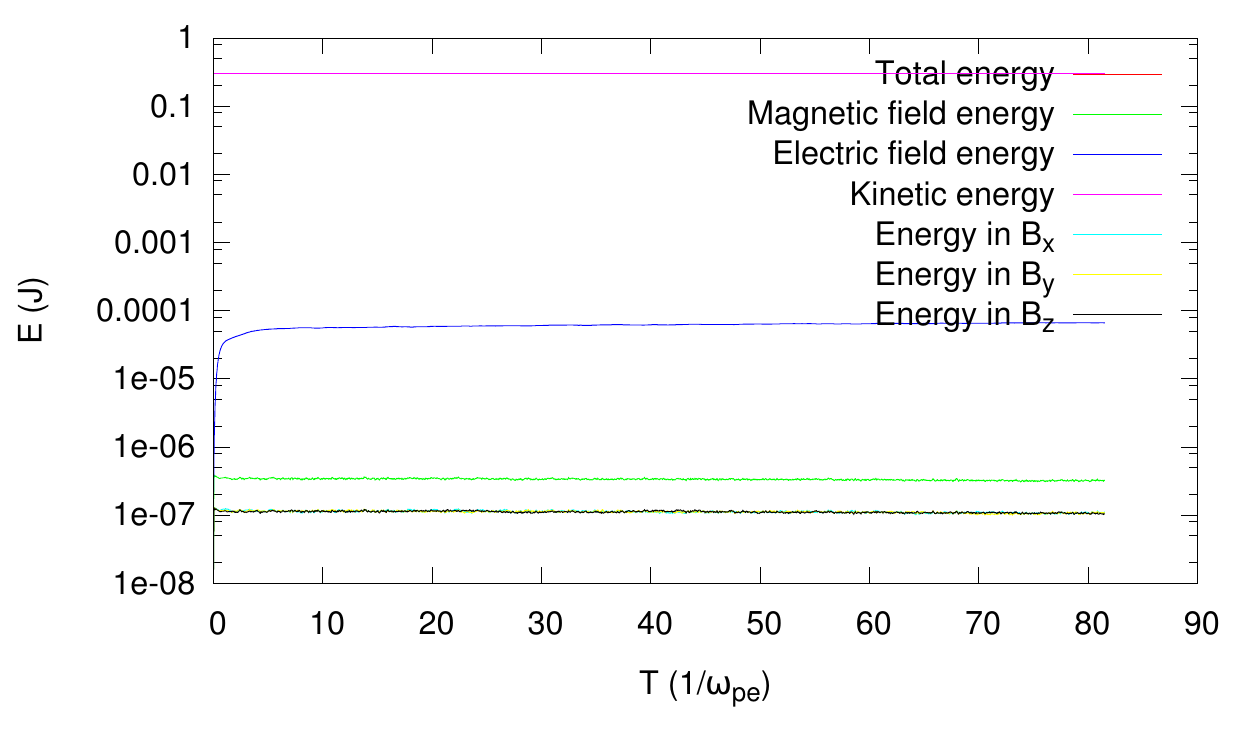}
		\includegraphics[width=0.48\textwidth]{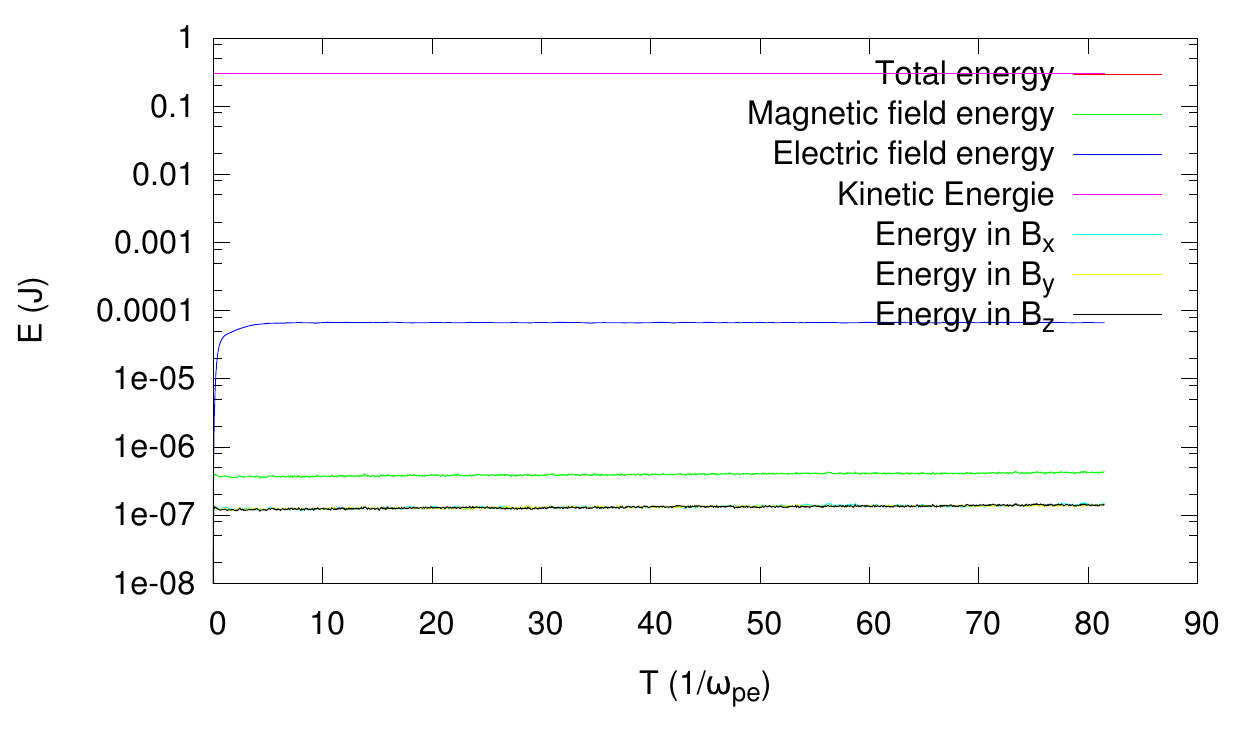}
		\includegraphics[width=0.48\textwidth]{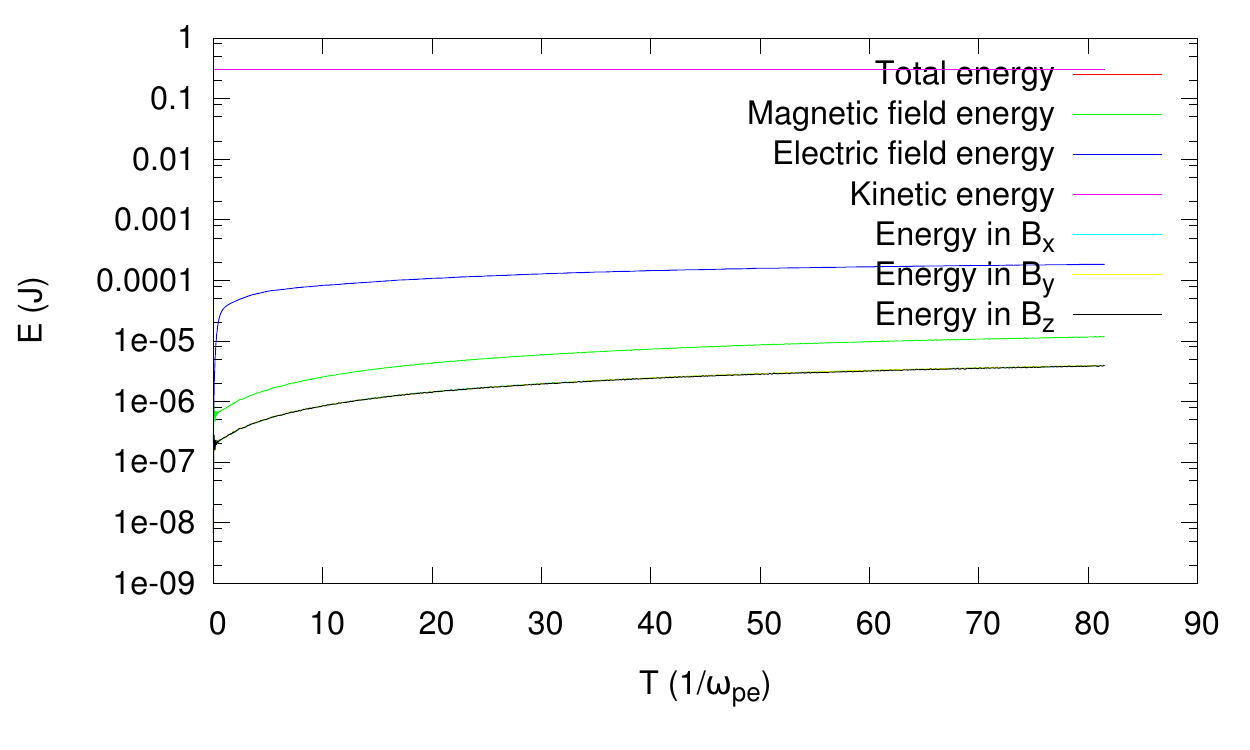}
		\caption{Energy development of the rescaled thermal simulation for the
		altered implicit scheme (top left), the explicit scheme used in ACRONYM
		(top right) and the original scheme (bottom).}
		\label{fig:energy}
	\end{figure}

	Finally, the simulations are repeated with $\Delta x$ and $\Delta t$ scaled
	down by a factor of 4, permitting a comparison with our explicit PiC Code
	ACRONYM \citep{Kilian_2011}. The energy development for the implicit scheme
	using the Yee-lattice, the explicit scheme, and the unaltered implicit
	scheme are shown in figure \ref{fig:energy}. The original scheme shows
	deviations from the explicit results with the magnetic field energy growing
	steadily over the course of the simulation. In contrast, the altered scheme
	nicely reproduces the energy development of the ACRONYM simulation.

	As shown, it is possible to alter the algorithm proposed by \citet{petdav}
	to obtain a divergence-free setup. In our code, the only steps that needed to
	be modified were the deposition and interpolation of grid quantities and the
	calculation of the updated fields. The former change is easily implemented by
	adjusting the offsets used when evaluating the weighting function. The latter
	change is described in section \ref{sec:modification}.

	Due to the additional calculations introduced into the scheme, some
	performance is lost. However, since most of the changes are confined to the
	field update step which only needs to be executed once per cell and timestep,
	the performance loss is small and on the order of a few percent.





\bibliographystyle{model1-num-names}
\bibliography{paper}







\end{document}